\documentclass[journal]{IEEEtran}

\usepackage{multirow}
\usepackage[table,xcdraw]{xcolor}
\usepackage{graphicx}
\usepackage{amsmath}
\usepackage{amsfonts}
\usepackage{algorithm}
\usepackage{amsthm,amssymb}
\usepackage{mathrsfs}
\usepackage{algorithmic}
\usepackage{makecell}
\usepackage{easyReview}
\usepackage{booktabs}
\setcellgapes{3pt}
\usepackage{verbatim}
\setcellgapes{3pt}

\usepackage[justification=centering]{caption}
\usepackage{color}
\usepackage[subfigure]{tocloft}
\usepackage{subfigure}
\usepackage{tabularx}
\usepackage{amsmath, bm}
\usepackage{hyperref}
\hypersetup{hidelinks,
	colorlinks=true,
	allcolors=blue,
	pdfstartview=Fit,
	breaklinks=true}

\usepackage{cite}

\ifCLASSINFOpdf
\else
\fi
\begin{document}
	\title{A Unified Cloud-Edge-Terminal Framework for Multimodal Integrated Sensing and Communication}
	\author{Yubo Peng, \textit{Graduated Student Member, IEEE}, Luping Xiang, \textit{Senior Member, IEEE}, Kun Yang, \textit{Fellow, IEEE}, Feibo Jiang, \textit{Senior Member, IEEE}, Kezhi Wang, \textit{Senior Member, IEEE}, and Christos Masouros, \textit{Fellow, IEEE}
		\thanks{
			{Yubo Peng (ybpeng@smail.nju.edu.cn), Luping Xiang (luping.xiang@nju.edu.cn), and Kun Yang (kunyang@nju.edu.cn) are with the State Key Laboratory of Novel Software Technology, Nanjing University, Nanjing, 210008, China, Institute of Intelligent Networks and Communications (NINE), and the School of Intelligent Software and Engineering, Nanjing University (Suzhou Campus), Suzhou, 215163, China.
    
            Feibo Jiang (jiangfb@hunnu.edu.cn) is with the School of Information Science and Engineering, Hunan Normal University, Changsha, China. 

            Kezhi Wang (Kezhi.Wang@brunel.ac.uk) is with the Department of Computer Science, Brunel University London, UK.

            Christos Masouros (c.masouros@ucl.ac.uk) is with the Department of Electronic and Electrical Engineering, University College London, London, UK.
			}
		}
	}

\markboth{Submitted for Review}%
{Shell \MakeLowercase{\textit{et al.}}: Bare Demo of IEEEtran.cls for IEEE Journals}

\maketitle 

\begin{abstract}
The transition to 6G calls for tightly integrated sensing and communication to support mission-critical services such as autonomous driving, embodied AI, and high-precision telemedicine. However, most existing ISAC designs rely on a single sensing modality (often RF), which limits environmental understanding and becomes a bottleneck in complex and dynamic scenes. This motivates a shift from single-modal to multimodal ISAC, where heterogeneous sensors (e.g., radar, LiDAR, and cameras) complement each other to improve robustness and semantic awareness.
In this article, we first summarize key challenges for multimodal ISAC, including heterogeneous fusion, communication overhead, and scalable system design. We then highlight three enabling technologies: large AI models, semantic communications, and multi-agent systems, and discuss how their combination can enable task-oriented multimodal perception. Building on these insights, we propose a unified cloud–edge–terminal (CET) framework that hierarchically distributes intelligence and supports three adaptive operation modes: global fusion mode (GFM), cooperative relay mode (CRM), and peer interaction mode (PIM). A case study evaluates the framework across three modes, demonstrating that GFM achieves the highest accuracy, PIM minimizes latency, and CRM strikes an optimal balance between performance and efficiency. Finally, we conclude with open research issues and future directions.
\end{abstract}


%
\IEEEpeerreviewmaketitle

\section{Introduction}
\subsection{Background}
Next-generation intelligent applications, such as autonomous driving, embodied artificial intelligence (AI), and high-precision telemedicine, inherently rely on the comprehensive environmental understanding and ultra-reliable connectivity \cite{10829757}. 
For instance, embodied robots require visual semantic understanding for navigation, whereas remote surgical manipulators demand haptic and visual feedback coupled with low-latency control. This underscores the critical role of integrated sensing and communication (ISAC) in future network architectures.

Although traditional ISAC systems effectively unify sensing and communication within a shared radio frequency (RF) framework to enhance spectral efficiency, relying exclusively on RF signals as the sole sensing modality often results in significant performance bottlenecks \cite{peng2025simac}.
While RF signals are proficient in range and velocity estimation, they offer a restricted single-modality perspective that is often insufficient for capturing the fine-grained semantic and contextual details demanded by complex environments.
Moreover, RF-based sensing is susceptible to severe performance degradation in highly dynamic scenarios caused by multipath propagation, interference, and occlusions.

These limitations necessitate a fundamental paradigm shift from single-modal to multimodal ISAC. 
In contrast to single-modal approaches, multimodal ISAC integrates data from diverse sensors operating on distinct physical principles, such as radar, Light Detection and Ranging (LiDAR), RGB-D cameras, and Global Positioning System (GPS), to construct a holistic representation of the physical world.
By leveraging the complementary strengths of these diverse modalities, multimodal ISAC offers significant potential for enhancing sensing accuracy, robustness, and generalization capabilities \cite{10330577}.
Existing research has already demonstrated the efficacy of this synergy; for instance, incorporating visual or spatial modalities can significantly accelerate the beam selection process. 
Notable implementations include LiDAR-aided \cite{8642397}, camera-GPS-aided \cite{10000718}, and camera-aided beam prediction \cite{10283784}, where multimodal data mitigates the latency associated with the exhaustive search protocols typical of conventional RF methods.
Motivated by its strategic importance in 6G and its distinct advantages over legacy systems, this article systematically explores the challenges, enabling technologies, and architectural frameworks necessary to realize intelligent multimodal ISAC.

\subsection{Contributions}
Motivated by the limitations of existing single-modal ISAC systems and the unique demands of 6G applications, this article presents several key contributions to promote the design and implementation of efficient and intelligent multimodal ISAC architectures, as follows:
\begin{enumerate}
   \item \textit{Challenge for multimodal ISAC:} We conduct a comprehensive analysis of the fundamental challenges in enabling multimodal ISAC. These challenges include (i) the effective fusion of heterogeneous data streams from diverse sensing modalities, (ii) the substantial communication overhead incurred by distributed information exchange among sensors, and (iii) the design of efficient, scalable, and adaptable system architectures capable of supporting real-time sensing and communication under dynamic network conditions.

    \item \textit{Technologies for multimodal ISAC:} We identify and critically examine several promising technologies that serve as enablers for multimodal ISAC, including large AI models (LAMs) \cite{10570717}, semantic communication (SC) \cite{10558819}, and multi-agent systems (MAS) \cite{10208208}. We discuss their complementary strengths, such as LAMs’ capacity for multimodal fusion, SC’s ability to task-oriented efficient transmission, and MAS’s support for distributed devices, and illustrate how their integration can significantly enhance the semantic understanding and coordination capabilities of multimodal ISAC systems.

    \item \textit{Unified cloud-edge-terminal Framework:} 
    We present a unified Cloud-Edge-Terminal (CET) framework that distributes intelligence hierarchically. Specifically, we design three dynamic operational modes: (i) Global Fusion Mode (GFM), utilizing vertical links for high-precision centralized fusion; (ii) Cooperative Relay Mode (CRM), employing edge nodes as semantic bridges for collaborative guidance; and (iii) Peer Interaction Mode (PIM), facilitating horizontal interactions for agile autonomy. This sequence reflects a continuum of progressive decentralization, effectively shifting the computational locus from the global cloud to the cooperative edge, and finally to the autonomous terminal.

    \item \textit{Case Validation:} We validate the proposed CET framework through a comprehensive case study on the DeepSense 6G dataset \cite{10144504}. The experimental results quantify the distinct advantages of each mode: the GFM maximizes accuracy, the PIM minimizes latency (offering a $\sim7\times$ speedup), and the CRM effectively bridges the gap, offering the best trade-off between performance and latency. This shows the necessity of a mode-adaptive framework to meet the requirements in different scenarios.
\end{enumerate}

\section{Challenges for Multimodal ISAC}
Despite its considerable potential to enhance environmental perception, the practical realization of multimodal ISAC faces several critical challenges stemming from the heterogeneous nature of sensing modalities and the complexity of real-time communication and computation. These challenges must be carefully addressed to ensure effective system design and deployment across diverse application scenarios.

\subsection{Heterogeneous Multimodal Fusion}
Multimodal ISAC inherently involves the integration of data streams originating from diverse sensing modalities such as RF signals, vision sensors, and LiDAR. These modalities differ substantially in terms of data dimensionality, spatial-temporal resolution, coverage, and semantic abstraction. For instance, aligning 2D visual images with 1D radar waveforms is particularly challenging due to discrepancies in data structure, coordinate systems, and information content \cite{10558825}. Moreover, inconsistencies in sensing frequency and perception latency further complicate synchronized fusion. 

\subsection{High Communication Overhead}
Real-time sensing and decision-making in distributed ISAC systems require frequent and high-volume data exchanges between sensors, edge nodes, and centralized servers \cite{11175176}. This results in considerable bandwidth consumption and elevated energy expenditure, particularly in wireless or resource-constrained environments. The issue becomes more pronounced in low-dynamic scenarios, such as static nighttime surveillance, where redundant or minimally informative data (e.g., successive identical video frames) continue to be transmitted, thereby wasting transmission resources without improving situational awareness. 

\subsection{Context-Aware Architectural Design}
Designing a system architecture that can adapt to diverse operational environments and task requirements remains a fundamental challenge. Rigid deployment modes may offer simplicity in control and resource management, but often lack the flexibility needed to accommodate the heterogeneous demands of ISAC applications. For example, while centralized architectures benefit from powerful computing infrastructure, they introduce latency and dependency issues in time-critical or infrastructure-sparse settings. Conversely, purely distributed systems offer autonomy but often suffer from limited coordination and reduced global situational awareness \cite{10979960}. 

Addressing these challenges calls for more intelligent context-aware systems, which is an area where several advanced technologies, such as LAMs, SC, and MAS, offer promising avenues for enabling adaptive multimodal fusion, communication-efficient sensing, and scalable architectural designs.

\begin{figure*}[htbp]
	\centering
	\includegraphics[width=17cm]{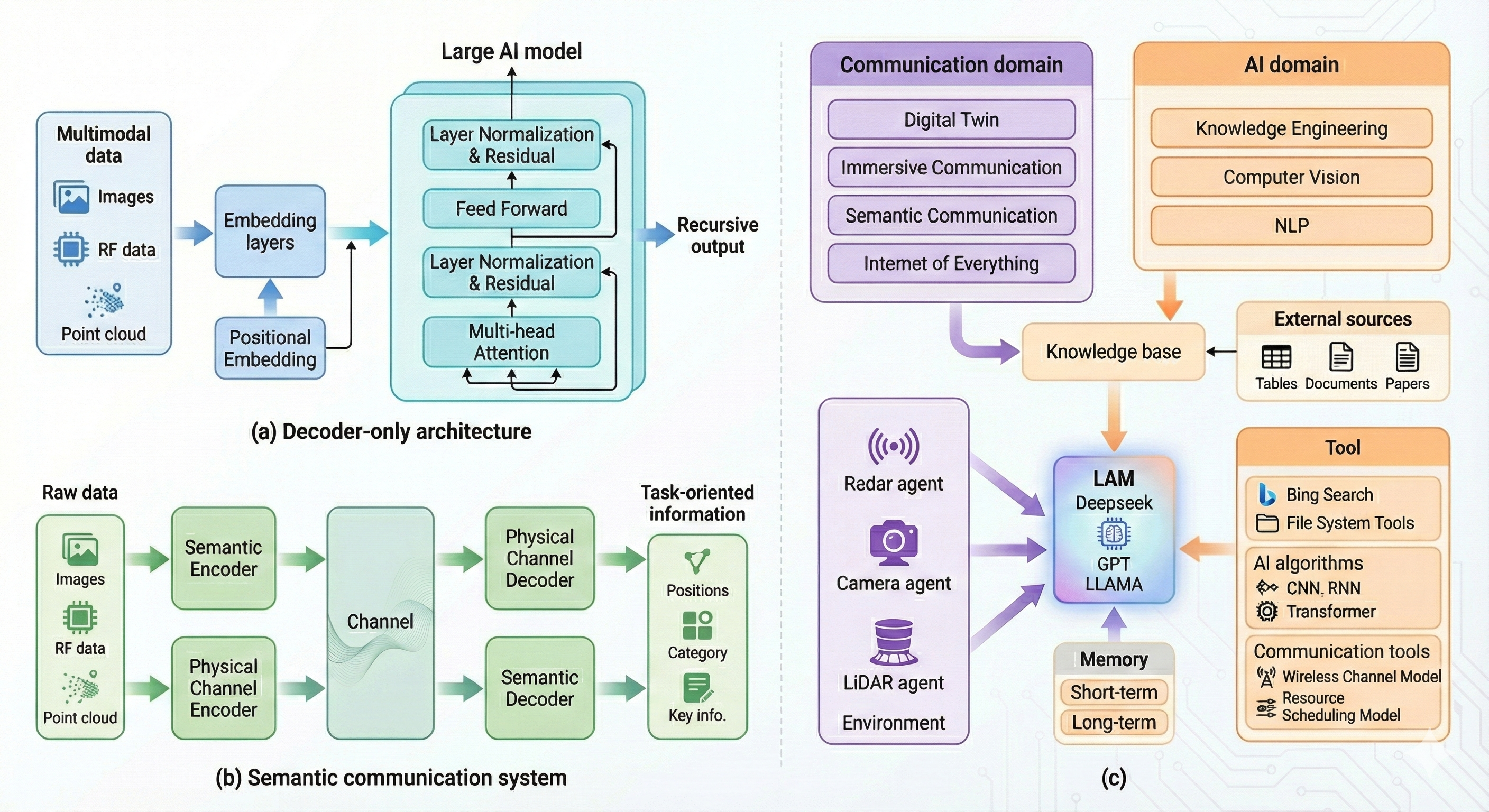}
	\caption{The illustration of three technologies. (a) Large AI models. (b) Semantic communication. (c) Multi-agent system.}
	\label{fig:tec}
\end{figure*}
\section{Technologies for Multimodal ISAC}
As shown in Fig. \ref{fig:tec}, this section introduces several key technologies, detailing how they address the challenges associated with realizing multimodal ISAC as previously discussed.

\subsection{Large AI Model}
LAMs demonstrate exceptional capabilities in processing heterogeneous multimodal data by leveraging their extensive parameter space and rich prior knowledge \cite{10570717}. As illustrated in Fig. \ref{fig:tec}(a), inputs from various modalities, such as images, RF signals, and point clouds, are first transformed into token representations through embedding layers. These tokens, combined with learnable positional embeddings, are then fed into a decoder-only transformer architecture. This architecture consists of stacked decoder blocks, each comprising masked multi-head self-attention for autoregressive modeling, followed by feed-forward networks, residual connections, and layer normalization. This token-based representation and fusion strategy enables LAMs to effectively extract and integrate both low-level physical features and high-level semantic information, thereby enhancing performance in tasks such as object recognition and motion estimation.

\subsection{Semantic Communication}
As illustrated in Fig. \ref{fig:tec}(b), a typical SC system comprises semantic and channel encoders at the transmitter, and corresponding channel and semantic decoders at the receiver \cite{10558819}. Unlike conventional communication systems that prioritize bit-level fidelity, SC first employs AI-based semantic encoders to extract essential semantics from raw multimodal data, including images, RF signals, and point clouds. Then, only the extracted semantic representations are transmitted, significantly reducing data volume. Finally, at the receiver, AI-based semantic decoders reconstruct task-specific information from the received semantics, such as object categories, positions, etc. 
This strategy effectively filters out redundant or irrelevant content, enabling efficient operation under stringent bandwidth constraints. 

\subsection{Multi-Agent System}
MAS \cite{10208208} provides the structural foundation for enabling scalable, decentralized intelligence within the multimodal ISAC. In MAS, multiple autonomous agents, either software-based or embodied in hardware, interact with the environment, as well as each other, to collaboratively achieve task objectives. Each agent perceives local conditions, makes decisions, and executes actions, often through cooperative reasoning and intent-aware communication.
As shown in Fig. \ref{fig:tec}(c), the core component of MAS includes four functional modules: a knowledge base for storing domain knowledge, a memory module for accumulating interaction history, a LAM, such as Deepseek and GPT, for reasoning and decision-making, and external tools for executing actions. During operation, an agent continuously interacts with the environment, stores relevant observations in memory, combines them with contextual knowledge from the knowledge base, and inputs the fused information into the LAM. The LAM then analyzes the situation and invokes appropriate tools (e.g., an AI or generic algorithm) to perform task-specific actions.

Overall, LAM offers robust capabilities for multimodal data processing, analysis, and fusion, as well as for global control and decision-making across heterogeneous sensors. SC ensures high-efficiency and low-overhead data transmission. MAS provides the organizational strategy, endowing individual sensors with intelligence to enable autonomous reasoning and collaborative behavior. 
While each of these technologies offers distinct advantages, determining how to effectively and appropriately deploy them in varying scenarios remains a critical challenge.

\begin{figure*}[htbp]
	\centering
	\includegraphics[width=17cm]{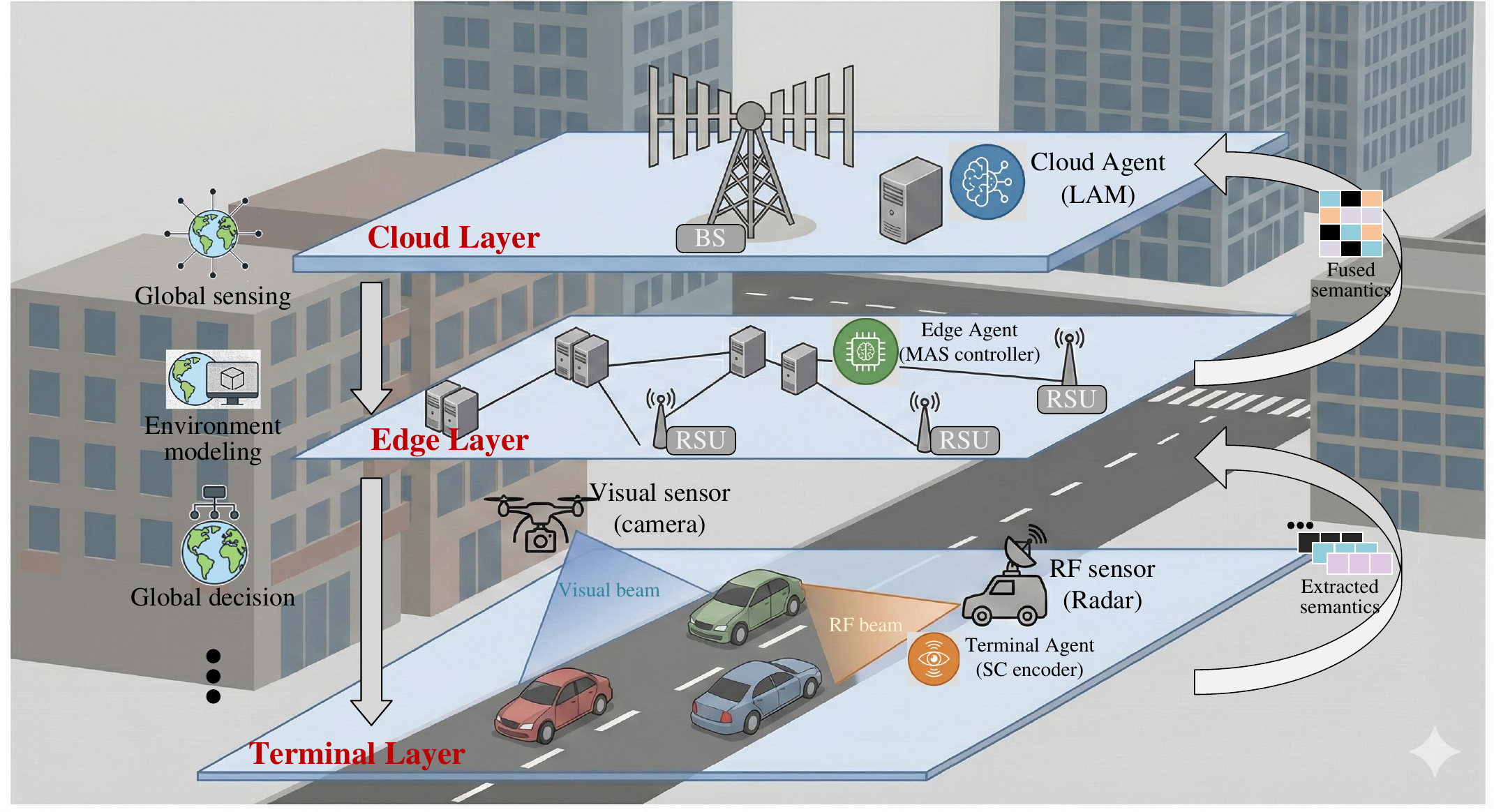}
	\caption{The illustration of the proposed CET framework for multimodal ISAC.}
	\label{fig:arch}
\end{figure*}

\section{Cloud-edge-terminal Collaborative Multimodal ISAC Architectures}
To fully harness the potential of the aforementioned enabling technologies (e.g., LAM, SC, and MAS) in the multimodal ISAC system, we propose a unified CET framework, as illustrated in Fig. \ref{fig:arch}. This hierarchical design is motivated by the practical constraints of emerging 6G applications, such as autonomous driving and urban surveillance. In real-world deployments, individual terminal sensors often lack the computational power to run LAMs locally, while relying solely on centralized cloud processing introduces prohibitive latency and excessive backhaul traffic for high-volume multimodal data. By distributing intelligence across three layers, this architecture strikes a balance between perception accuracy and resource efficiency. 
The terminal layer consists of heterogeneous sensors (e.g., cameras and radars) equipped with lightweight SC encoders for initial semantic extraction. The edge layer hosts MAS controllers at roadside units or base stations to manage local coordination and low-latency task offloading. Finally, the cloud layer leverages the powerful LAM to maintain global environment models and perform deep reasoning. 

\begin{figure*}[htbp]
	\centering
	\includegraphics[width=18cm]{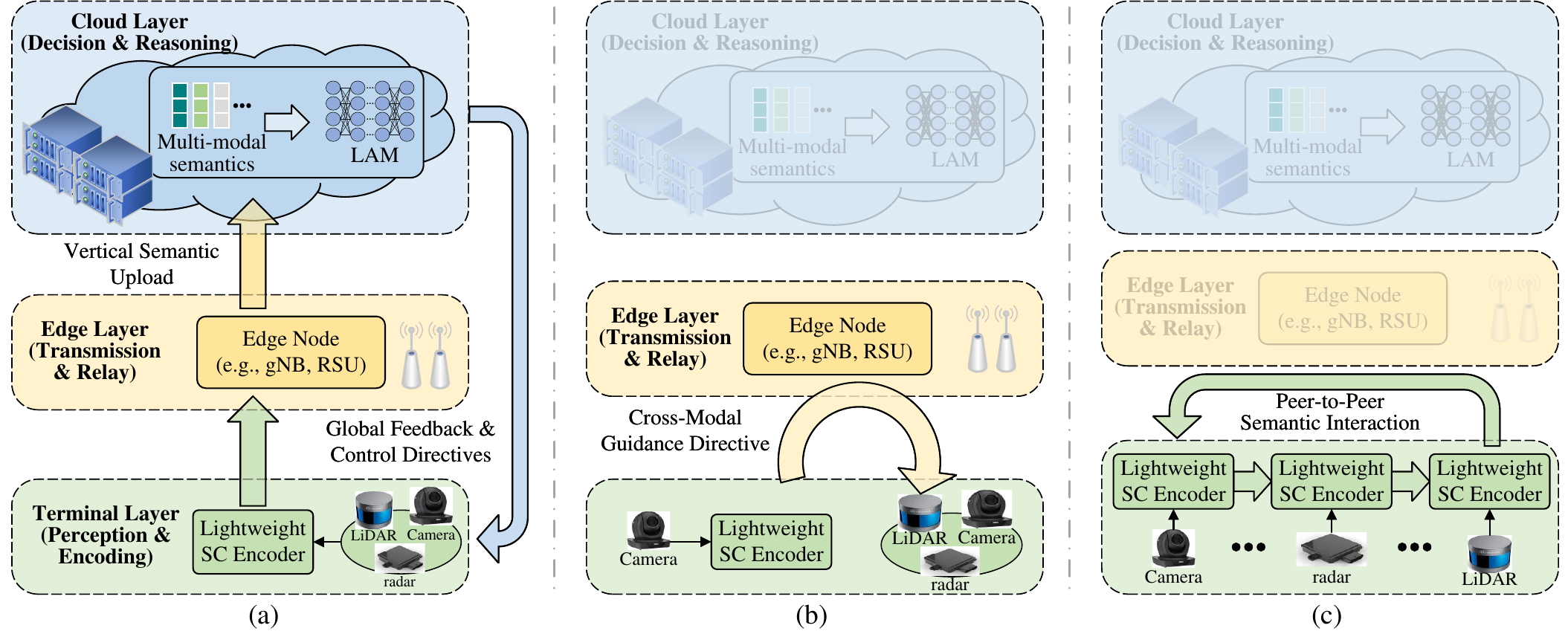}
	\caption{The illustration of the three modes in the CET framework.(a) GFM, (b) CRM, and (c) PIM.}
	\label{fig:mode}
\end{figure*}

To accommodate diverse sensing tasks and dynamic network conditions, as shown in Fig. \ref{fig:mode}, the proposed framework operates in three distinct modes: GFM, CRM, and PIM. 
This sequence represents a progressive decentralization of intelligence, shifting the computational locus from the global cloud (GFM) to the cooperative edge (CRM), and finally to the autonomous terminal (PIM). The system dynamically transitions between these modes to optimize the trade-off between sensing accuracy, latency, and coverage.

\subsection{Global Fusion Mode}
As illustrated in Fig. \ref{fig:mode}(a), the GFM corresponds to the vertical data flow within our collaborative architecture, adopting a centralized fusion strategy to integrate semantic information from heterogeneous sensors. 
Specifically, multimodal sensors at the terminal layer are equipped with lightweight edge agents. These agents serve as semantic encoders, extracting and compressing task-relevant semantics from local multimodal data. 
Then, the encoded compact semantic representations are transmitted vertically to the cloud layer over wireless channels, passing through the edge infrastructure. 
At the cloud layer, a powerful LAM serves as the central agent to align and fuse the received multimodal semantics to generate unified global representations. 
The fused features are then utilized for downstream applications such as global decision-making, collaborative sensing, and large-scale environmental modeling. 

A typical application scenario for GFM is intelligent transportation, where cameras, LiDAR sensors, and RF devices are distributed across various intersections \cite{9830717}. 
In this context, each terminal sensor, via its corresponding edge agent, extracts semantic information such as vehicle positions, motion trajectories, and Doppler shifts. 
These compressed semantic features are aggregated at a central LAM located at a cloud-based traffic control center. 
Here, global fusion is performed to achieve a comprehensive understanding of real-time traffic conditions across the entire city grid, thereby enabling intelligent traffic signal coordination and city-wide congestion mitigation.

In the GFM, the deployment of a powerful LAM at the central cloud enables deep reasoning over globally aggregated data, significantly improving decision accuracy. 
Meanwhile, SC offloads bandwidth-intensive raw data transmission from the terminal to the cloud, ensuring efficiency even in bandwidth-limited environments. 
However, this mode relies heavily on the connectivity to the Cloud Layer and the computational availability of the central agent. 
Consequently, when communication links between terminal sensors and the cloud are disrupted, or when latency constraints are stringent, the system may suffer from reduced responsiveness. 
This dependency on centralized infrastructure necessitates a shift towards edge-assisted coordination, leading to the Cooperative Relay Mode.

\subsection{Cooperative Relay Mode}
Fig. \ref{fig:mode}(b) illustrates the CRM, which serves as the intelligent bridging mechanism within our architecture, shifting the operational focus from the cloud to the edge layer.
In this mode, the edge layer (e.g., roadside units (RSUs)) acts as a semantic relay and coordinator between heterogeneous sensors at the terminal layer. 
Unlike the vertical upload in GFM, CRM utilizes the edge node as an intermediary to facilitate local cross-modal coordination.
Terminal sensors equipped with moderate edge agents extract essential feature semantics and transmit them to a nearby relay node. 
The relay agent then interprets, repackages, and forwards these semantics, often translating them into control directives, to other terminal agents operating on different modalities. 
This enables a ``sense-and-guide" workflow where one modality cues another without burdening the central cloud.

A prime example of CRM is the drone-assisted surveillance scenario. 
Here, a camera-equipped drone (Terminal A) employs its onboard agent to extract positional semantics from aerial observations. 
This semantic information is transmitted to an RSU-based relay agent. 
Referencing a lightweight knowledge base, the relay agent formulates a precise semantic directive (e.g., ``focus radar on coordinates (x, y)") and forwards it to a ground-based radar sensor (Terminal B). 
Upon receiving the directive, the radar’s agent dynamically adjusts its ISAC beam toward the specified coordinates. 
This cooperative mechanism avoids exhaustive scanning and significantly improves sensing efficiency by leveraging the wide field-of-view of vision to guide the precise ranging capabilities of RF.

Compared to the GFM, CRM reduces dependency on the central cloud and supports extended spatial coverage through multi-hop relaying via edge nodes. 
However, it still relies on an infrastructure-based intermediary (the relay) to perform semantic translation and coordination.
In scenarios demanding ultra-low latency or where infrastructure is unavailable, even the delay introduced by edge relaying may be unacceptable. This drives the need for a direct, infrastructure-free approach, as realized in the Peer Interaction Mode.

\begin{table*}[htbp]
\centering
\renewcommand{\arraystretch}{1.3}
\caption{Comparison of Three Architectural Schemes}
\label{tab:cap}
\begin{tabularx}{\textwidth}{|>{\centering\arraybackslash}X
                             |>{\centering\arraybackslash}X
                             |>{\centering\arraybackslash}X
                             |>{\centering\arraybackslash}X|}
\hline
\textbf{Aspect} & \textbf{GFM (Cloud-based)} & \textbf{CRM (Edge-based)} & \textbf{PIM (Terminal-based)} \\
\hline
\textbf{Sensing Coverage} & 
Global / Macro-scale & 
Regional / Zone-based & 
Local Neighborhood / One-hop \\
\hline
\textbf{Strengths} & 
High-accuracy inference; Strong multimodal fusion & 
Balanced workload; Moderate communication efficiency & 
High autonomy; Strong scalability; Robust resilience \\
\hline
\textbf{Weaknesses} & 
Strong central dependency; Low fault tolerance & 
Protocol dependency; Compromised global/local reasoning & 
Cognitive inconsistency; High terminal computational burden \\
\hline
\textbf{Communication Load} & 
High & 
Moderate & 
Low \\
\hline
\textbf{Use Cases} & 
Intelligent transportation hubs & 
Drone surveillance & 
Industrial inspection sites \\
\hline
\end{tabularx}
\end{table*}

\subsection{Peer Interaction Mode}
Fig. \ref{fig:mode}(c) depicts the PIM, which represents the final stage of decentralization: the horizontal cooperative path solely within the terminal layer. 
Designed to maximize autonomy, this mode enables direct, peer-to-peer (P2P) semantic interaction among distributed sensors, bypassing both the central cloud and edge relays.
Each terminal agent, driven by lightweight language models, independently processes local multimodal inputs and transmits only essential high-level semantic outcomes, such as alerts, decisions, or event summaries, directly to neighboring terminal agents.
Subsequently, receiving agents perform local analysis and reasoning by leveraging this shared semantic context, internal memory, and localized knowledge bases.

A representative application scenario for PIM is industrial inspection, where cameras, vibration sensors, and acoustic sensors are deployed along a production line \cite{10086648}. 
When a vibration sensor agent detects an anomaly and a camera agent simultaneously identifies a visual defect, they can directly exchange semantic messages via the horizontal link to collaboratively diagnose and localize faults in real time. 
This direct terminal-to-terminal collaboration eliminates the round-trip latency to the cloud or edge, ensuring rapid response times critical for safety-sensitive tasks.

By removing the reliance on upper-layer infrastructure (cloud or edge), PIM achieves the highest level of resilience and is particularly effective in bandwidth-constrained or infrastructure-deficient environments. 
Nonetheless, the absence of a centralizing brain (GFM) or a coordinating relay (CRM) introduces challenges regarding cognitive consistency. 
Data heterogeneity and differences in prior knowledge across distributed terminal agents may lead to divergent interpretations. 
For example, without a higher-level node to align discrepancies, a camera-based agent and an infrared-based agent may generate conflicting local decisions under varying environmental conditions.
These limitations highlight that while PIM offers superior speed and autonomy, dynamic switching back to CRM or GFM is necessary when higher-order reasoning or global consensus is required.

To clearly illustrate the differences among the three modes, Table \ref{tab:cap} provides a comparative summary of their key characteristics.

\section{Case Study}

\subsection{Experimental Settings}
\subsubsection{Data Settings}
To evaluate the proposed CET framework under realistic multimodal ISAC conditions, we utilize the DeepSense 6G dataset \cite{10144504}, specifically focusing on Scenario 31 (daytime street) and Scenario 33 (nighttime street). These scenarios offer a diverse range of environmental conditions essential for testing robustness. We implement a unified data loading pipeline that supports batched, shuffled, and parallel sampling. To mitigate data imbalance across heterogeneous modalities, mini-batches are sampled from scenario-specific subsets using a probability distribution proportional to the dataset size, ensuring balanced training dynamics. Data preprocessing follows standard protocols in multimodal sensing to guarantee numerical stability:
(i) RGB images: Resizing, normalization, and standard augmentation techniques (e.g., random cropping) are applied.
(ii) RF power: Sequences undergo min--max normalization and NaN value imputation.
(iii) LiDAR point clouds: Data is centered, scaled to a unit sphere, and padded to a fixed point count.
(iv) Radar mmWave: Complex-valued IQ signals are normalized and reshaped into consistent sequence formats.

\subsubsection{Baseline Settings}
To validate the efficacy of the proposed CET framework, we benchmark against three distinct operation modes. The primary differentiator is the number of accessible sensing modalities, determined by the mode's coverage capabilities (as illustrated in Table \ref{tab:cap}).

\begin{itemize}
    \item GFM: This represents the full collective mode of cloud-edge-terminal. We assume a centralized cloud server with global coverage and powerful LAMs, enabling it to aggregate and process all available modalities from the terminal: Image (I), RF power (P), Point Cloud (C), and mmWave (M). This setup serves as the theoretical upper bound of sensing accuracy.
    
    \item CRM: This baseline represents the edge-terminal collective mode. We simulate a cooperative scenario involving nearby edge nodes (e.g., RSU or V2V pairs). Through short-range communication, the primary agent can augment its local sensing with an additional modality from a neighbor, expanding the accessible set to three modalities. This mode balances coverage and overhead. The specific variants are:
    \begin{enumerate}
        \item CRM (P+I+C): Fuses RF Power, RGB Image, and LiDAR.
        \item CRM (P+I+M): Fuses RF Power, RGB Image, and mmWave.
        \item CRM (P+C+M): Fuses RF Power, LiDAR, and mmWave.
    \end{enumerate}

    \item PIM: This represents the terminal mode. We simulate a standalone resource-constrained device (e.g., NVIDIA Jetson) relying solely on onboard sensors. Due to hardware limitations, it processes only dual-modality inputs. This setting evaluates performance under strict resource constraints. The variants include:
    \begin{enumerate}
        \item PIM (P+I): Fuses RF Power and RGB Images.
        \item PIM (P+C): Fuses RF Power and LiDAR Point Clouds.
        \item PIM (P+M): Fuses RF Power and mmWave Radar data.
    \end{enumerate}
\end{itemize}

This tiered experimental design aims to quantify the trade-offs between the comprehensive sensing of GFM, the collaborative enhancement of CRM, and the lightweight efficiency of PIM.

\subsubsection{Metric Settings}
We adopt a comprehensive evaluation strategy that balances communication performance with system efficiency. Two primary categories of metrics are considered:

\begin{itemize}
    \item Beam Prediction Accuracy: Formulated as a multi-class classification task, this metric evaluates the Top-1 accuracy of the predicted optimal beam index. It serves as a direct proxy for the communication gain (e.g., effective achievable rate) provided by each sensing scheme.
    
    \item Computational Complexity: To assess the deployment feasibility on resource-constrained edge and terminal devices, we analyze the system efficiency through three distinct indicators:
    \begin{enumerate}
        \item Floating Point Operations (FLOPs): This measures the theoretical computational cost of the model, reflecting the processing power required.
        \item Memory Footprint: We monitor the peak GPU memory (VRAM) usage during inference to evaluate the spatial complexity and hardware storage requirements.
        \item Inference Latency: This records the average end-to-end execution time per sample, which is critical for determining the real-time capability in dynamic ISAC scenarios.
    \end{enumerate}
\end{itemize}

\subsection{Evaluation Results}
\subsubsection{Performance Analysis}
\begin{figure}[htbp]
	\centering
	\includegraphics[width=8.5cm]{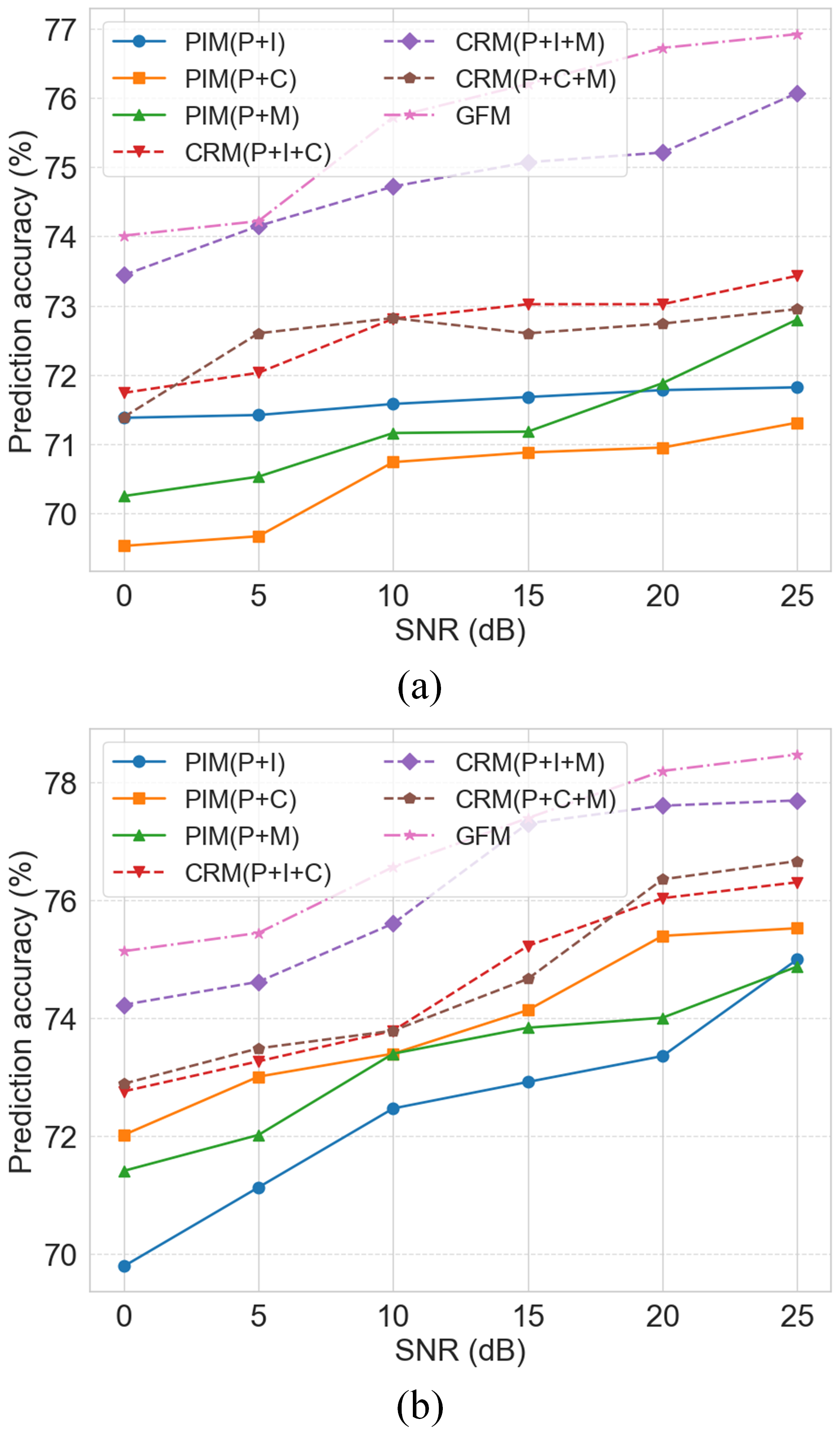}
	\caption{The beam prediction accuracy under different schemes in (a) Scenario 31 and (b) Scenario 33.}
	\label{fig:exp}
\end{figure}

The beam prediction accuracy versus Signal-to-Noise Ratio (SNR) for Scenario 31 (daytime) and Scenario 33 (nighttime) is illustrated in Fig. \ref{fig:exp}(a) and Fig. \ref{fig:exp}(b), respectively. 

It can be observed that the GFM consistently achieves the highest prediction accuracy across all SNR regimes in both scenarios. For instance, in Fig. \ref{fig:exp}(a), when SNR=25 dB, GFM reaches an accuracy of approximately 76.9\%, outperforming all other variants. This superiority is attributed to the GFM's ability to leverage the comprehensive information from more modalities simultaneously, effectively establishing a performance upper bound for the system.
The CRM effectively bridges the performance gap between the cloud and the terminal. As shown in Fig. \ref{fig:exp}(b), while the performance of the vision-dependent PIM (P+I) degrades significantly due to poor lighting conditions at night, the cooperative CRM (P+I+M) leverages the additional mmWave radar modality from a neighbor to maintain high accuracy, closely approaching the GFM. 
In contrast, the PIM exhibits varying degrees of limitation. This performance hierarchy (GFM $>$ CRM $>$ PIM) highlights the trade-off between modality richness and sensing accuracy.

\subsubsection{Complexity Analysis}
\begin{table}[htbp]
  \centering
  \caption{Computational complexity comparison}
  \label{tab:complexity}
  \begin{tabular*}{\linewidth}{@{\extracolsep{\fill}}lccc@{}}
    \toprule
    Method & FLOPs (G) & Memory (MB) & Inference Latency (ms) \\
    \midrule
    PIM(P+I)  & 4.15e+00 & 2.40e+01 & 7.27e+00 \\
    PIM(P+C)  & 9.45e+00 & 1.05e+02 & 3.22e+01 \\
    PIM(P+M)  & 7.30e+00 & 1.89e+01 & 5.51e+00 \\
    CRM(P+I+C) & 9.45e+00 & 1.07e+02 & 3.11e+01 \\
    CRM(P+I+M) & 7.31e+00 & 2.46e+01 & 1.33e+01 \\
    CRM(P+C+M) & 8.46e+00 & 1.06e+02 & 3.42e+01 \\
    GFM     & 1.26e+01 & 1.07e+02 & 3.82e+01 \\
    \bottomrule
  \end{tabular*}
\end{table}

Table \ref{tab:complexity} presents the computational complexity per sample in terms of FLOPs, memory footprint, and inference latency (excluding transmission delay).

In terms of resource consumption, the GFM imposes a substantial burden. While this high complexity underpins its superior accuracy, it represents an approximately $3\times$ increase in computation compared to the lightweight PIM (P+I). Such heavy resource demands make the GFM challenging to deploy directly on power-constrained edge devices.
Regarding execution speed, the PIM variants demonstrate significant advantages, operating with extremely low inference latency. Notably, the CRM occupies a strategic ``sweet spot." This indicates that CRM can support sophisticated multi-modal fusion without the heavy latency penalty of the full cloud model.
Moreover, the practical deployment latency extends beyond computation to include data transmission overhead:
\begin{itemize}
    \item \textit{GFM:} Necessitates uploading the feature data from the terminal to the cloud via edge nodes. Assuming a moderate uplink bandwidth of 50 Mbps, the transmission latency could exceed 300 ms, making the total system latency ($>340$ ms) unsuitable for delay-sensitive beam tracking.
    \item \textit{CRM:} Relies on high-bandwidth, short-range D2D/V2X links (e.g., neighbors). The transmission latency is significantly lower than the cloud uplink, ensuring real-time responsiveness.
    \item \textit{PIM:} Processes data locally, offering the lowest possible total latency ($\sim$ 5-10 ms).
\end{itemize}

The experimental results validate the motivation of our proposed CET framework. While GFM provides the accuracy upper bound and PIM ensures the latency lower bound, the CRM offers a flexible balance. This necessitates a collaborative architecture that dynamically switches between these modes to satisfy the conflicting requirements of high-precision and low-latency ISAC.

\section{Open Issues and Future Directions}
In the proposed CET framework, although the integration of LAMs, SC, and MAS enhances sensing efficiency through distributed intelligence, expands the attack surface across the cloud, edge, and terminal layers. These new security challenges transcend traditional physical-layer jamming.

\subsection{Semantic Tampering in Global Fusion}
In the GFM, terminal agents compress raw data into semantic features for uplink transmission to the cloud. Unlike traditional bit-flipping attacks, adversaries may launch semantic tampering attacks by perturbing the feature vectors in the latent space. Such attacks can impact high-level interpretations (e.g., changing a ``pedestrian" label to ``background") without disrupting the communication link or triggering conventional error correction mechanisms, thereby poisoning the global environmental model at the cloud. 
Therefore, we should explore new semantic integrity verification mechanisms, such as embedding fragile watermarks or cryptographic signatures directly into the semantic latent space, alongside training robust SC models via adversarial learning to resist feature perturbations.

\subsection{Malicious Relaying and Instruction Injection in Cooperative Edge}
The CRM relies on edge nodes to act as semantic relays, translating observations into control directives for other terminals (e.g., guiding a radar beam based on visual inputs). This may make the system vulnerable to malicious relaying and instruction injection. A compromised edge node could alter the semantic intent of a message or inject fake guidance commands, effectively blinding the sensing system or forcing terminals to waste resources scanning space.
To mitigate this, a zero-trust architecture is required where terminal agents cryptographically verify the legitimacy of edge directives. Additionally, behavior-based anomaly detection can be deployed to identify and isolate edge nodes that issue statistically improbable or contradictory control commands.

\subsection{Cross-Modal Misleading in Peer Interaction}
In the PIM, autonomous terminals exchange high-level semantic alerts directly to facilitate rapid local collaboration. However, this decentralization is vulnerable to cross-modal misleading, where a compromised sensor (e.g., a camera) broadcasts falsified object detections to its peers. Without a central authority to verify ground truth, honest peer agents (e.g., radars) may be tricked into resolving cognitive conflicts in favor of the false data, leading to collective hallucinations.
As a solution, we can develop physics-aware consistency checks, where agents utilize physical laws (e.g., correlating Doppler shifts with visual optical flow) to validate peer data. Furthermore, a decentralized reputation management system can be established to dynamically reduce the trust weight of nodes that frequently report uncorroborated semantics.

\section{Conclusion}
In this article, we argued that multimodal ISAC is a necessary evolution beyond RF-only designs to meet the sensing and connectivity requirements of emerging 6G applications. After summarizing key challenges (heterogeneous fusion, communication overhead, and scalable architectures), we reviewed enabling technologies (LAM, SC, MAS) and proposed a CET framework with three adaptive modes (GFM/CRM/PIM). A case study demonstrated the unique advantages of different modes in this framework, and we outlined open issues for future research.

\bibliographystyle{IEEEtran}
\bibliography{bare_jrnl}




\newpage
\end{document}